\documentclass{article} % For LaTeX2e
\usepackage{times}
\usepackage{amsmath,amsfonts,bm}

\def\eqref#1{equation~\ref{#1}}
\def\1{\bm{1}}

\DeclareMathAlphabet{\mathsfit}{\encodingdefault}{\sfdefault}{m}{sl}
\SetMathAlphabet{\mathsfit}{bold}{\encodingdefault}{\sfdefault}{bx}{n}

\usepackage{hyperref}
\usepackage{url}
\usepackage{adjustbox}
\usepackage{pgfplots}
\usepackage{graphicx}
\usepackage{caption}
\usepackage{subcaption}
\usepackage{float}
\usepackage[many]{tcolorbox}
\usepackage{enumitem}

\setlist[itemize]{noitemsep}

\usepackage{comment}
\usepackage{algorithm}
\usepackage{algpseudocode}
\algrenewcommand\algorithmicrequire{\textbf{Input:}}
\algrenewcommand\algorithmicensure{\textbf{Output:}}

\pgfplotsset{compat=1.16}
\usetikzlibrary{pgfplots.colorbrewer,}
\pgfplotsset{
    cycle list/.define={my marks}{
        every mark/.append style={solid,fill=\pgfkeysvalueof{/pgfplots/mark list fill}},mark=*\\
        every mark/.append style={solid,fill=\pgfkeysvalueof{/pgfplots/mark list fill}},mark=square*\\
        every mark/.append style={solid,fill=\pgfkeysvalueof{/pgfplots/mark list fill}},mark=triangle*\\
        every mark/.append style={solid,fill=\pgfkeysvalueof{/pgfplots/mark list fill}},mark=diamond*\\
    },
}
\pgfplotsset{
  errorBars/.style={
    error bars/error bar style={very thick},
    error bars/error mark options={very thick,solid,mark size=3pt,rotate=90},
    error bars/y dir=both,
    error bars/y explicit,
  }
}
\def\addlegendimage{\csname pgfplots@addlegendimage\endcsname}
\usepgfplotslibrary{external}
\definecolor{main}{HTML}{5989cf}
\definecolor{sub}{HTML}{cde4ff}
\newtcolorbox{boxH}{
    sharpish corners, % better drop shadow
    boxrule = 0pt,
    toprule = 4.5pt, % top rule weight
}

\makeatletter
\let\@fnsymbol\@arabic
\makeatother

\title{Can DeepFake Speech be Reliably Detected?}

\author{Hongbin Liu$^\dagger{}^*$\thanks{Work done during internship at Google}, Youzheng Chen$^\ddagger$, Arun Narayanan$^\ddagger$, Athula Balachandran$^\ddagger$\\
Pedro J. Moreno$^\ddagger$, Lun Wang$^\ddagger{}^*$\\
$\dagger$ Duke University, $\ddagger$ Google LLC \\
\texttt{hongbin.liu@duke.edu,lunwang@google.com}
}

\date{}

\begin{document}

\maketitle
\def\thefootnote{*}\footnotetext{Equal Contribution}
\begin{abstract}
Recent advances in text-to-speech (TTS) systems, particularly those with voice cloning capabilities, have made voice impersonation readily accessible, raising ethical and legal concerns due to potential misuse for malicious activities like misinformation campaigns and fraud.  While synthetic speech detectors (SSDs) exist to combat this, they are vulnerable to ``test domain shift", exhibiting decreased performance when audio is altered through transcoding, playback, or background noise.  This vulnerability is further exacerbated by deliberate manipulation of synthetic speech aimed at deceiving detectors. This work presents the first systematic study of such active malicious attacks against state-of-the-art open-source SSDs. White-box attacks, black-box attacks, and their transferability are studied from both attack effectiveness and stealthiness, using both hardcoded metrics and human ratings. The results highlight the urgent need for more robust detection methods in the face of evolving adversarial threats.
\end{abstract}

\section{Introduction}
\label{sec:intro}

\newenvironment{thisnote}{\par\color{blue}}{\par}

Recent years have witnessed a remarkable advance in text-to-speech (TTS) systems~\cite{kreuk2022audiogen,borsos2023audiolm,leng2023prompttts,saeki2023virtuoso,shen2023naturalspeech,wang2023valle,chattts,chen2024valle2,coqui,elevenlabs,le2024voicebox,lux2024massive}.
Some of these systems possess the zero-shot / few-shot voice cloning capability~\cite{biadsy2024zero,cooper2020zero,casanova2022yourtts,ye2024flashspeech,le2024voicebox}, that can mimic someone's voice using only a brief sample of that person's speech recording.
The widespread access of these systems, either through open-source projects or commercial APIs, has made voice impersonation easier than ever. 
This raises significant ethical and legal concerns as the capability can be easily misused for misinformation campaign, fraud, copyright infringement, \emph{etc}.
As an instance, a scammer utilized synthetic audio to mimic President Biden in unlawful robocalls during a New Hampshire primary election, resulting in a \$6 million penalty and felony accusations~\cite{sixmillion}.
Besides, numerous instances of synthetic audio misuses can be found online if searching with ``deepfake audio'', which underscores the urgent need to address this growing problem.

Synthetic speech detectors (SSDs) are deployed to mitigate the misuse of synthetic speech.
While there are a number of advanced SSDs~\cite{rawnet2,rawnetgatst,aasist}, recent research~\cite{muller2022does,xie2024codecfake} suggests they might struggle when facing ``test domain shift''.
Concretely, a detector may exhibit decreased performance when presented with audio that has undergone alterations, including transcoding, playback, background noise, or even just a shift in the TTS system used.
However, current research efforts are directed towards these natural changes to audio, it is important to recognize that deliberate manipulation of synthetic speech by an attacker with the intent to deceive detectors can significantly increase the likelihood of success of the attacker.
However, there is a lack of systematic research on these malicious perturbations.

In this work, we conduct the first systematic study of active malicious attacks against the state-of-the-art open-source SSDs.
We investigate a range of attack scenarios, considering adversaries with varying levels of access to the target SSD: those with full knowledge of the model (white-box), those who can only interact with it and observe the results (black-box), and those who cannot even query the model (agnostic); and evaluate the results using both hard-coded metrics and human ratings.

Our findings reveal that:
\begin{itemize}
    \item Increased access to the detector makes it easier for attackers to create deepfakes that can evade detection without any noticeable loss in audio quality.
    \item Existing open-source SSD detectors are vulnerable when facing synthetic audio generated by TTS systems never seen during training.
    \item VisQOL scores and human ratings show that the audio quality after attack is reasonable.
    \item Alarmingly, even in the agnostic setting, attackers can still bypass state-of-the-art open-source SSDs with reasonable chance.
\end{itemize}
\textbf{Overall, we need more robust SSDs to mitigate the growing threat of deepfake audio misuse.}

\section{Preliminary}

\subsection{TTS Techniques}

TTS systems, which convert written text into speech, have a long history of development and have made remarkable progress in recent years.
Early TTS systems primarily used a concatenative approach~\cite{khan2016concatenative}, where speech was synthesized by joining pre-recorded speech units from a database.
Despite the simplicity, they suffered from unnatural prosody and robotic-sounding speech.
To address the issue, researchers proposed statistical parametric speech synthesis~\cite{zen2009statistical}.
These systems used statistical models to learn the relationship between linguistic features (e.g., phonemes, part-of-speech tags) and acoustic features (e.g., fundamental frequency, spectral envelope) and enabled more natural-sounding speech generation with improved prosody.

Recently, deep learning has revolutionized the field of TTS. Neural network-based TTS systems have surpassed traditional methods in terms of speech naturalness and intelligibility. Several architectures have been explored, including sequence-to-sequence models~\cite{wang2017tacotron,ping2017deep}, attention-based models~\cite{ren2019fastspeech,ren2020fastspeech2}, and generative adversarial networks (GANs)~\cite{kumar2019melgan}. These models can directly learn the mapping from text to speech, enabling end-to-end training and eliminating the need for complex feature engineering. Some of the popular neural TTS systems include \cite{kreuk2022audiogen,borsos2023audiolm,leng2023prompttts,saeki2023virtuoso,shen2023naturalspeech,wang2023valle,chattts,chen2024valle2,coqui,elevenlabs,le2024voicebox,lux2024massive}.

\subsection{Synthetic Speech Detection Techniques}
In the past, detecting synthetic speech required carefully crafted features~\cite{doddington2001speaker,alegre2013new,hanilcci2015classifiers,patel2015combining,sahidullah2015comparison,todisco2016new}.
However, with the increase in available data and the development of larger models, simpler features like waveforms or spectrograms are now sufficient for effective detection.
RawNet2~\cite{rawnet2} is a deep convolutional neural network for synthetic speech detection using merely raw waveforms.
It builds upon the RawNet~\cite{rawnet} architecture by incorporating residual connections and dilated convolutions.
RawNetGATST~\cite{rawnetgatst} extends RawNet2 by incorporating a graph attention network~\cite{tak2021graph} to identify key spectral or temporal features for detection.
Similarly, AASIST~\cite{aasist} refines the graph network architecture further for improved synthetic speech detection.

\section{Bypass Synthetic Speech Detection Systems}

In this section, we aim to answer the following question:

\noindent\textbf{Can deepfake audio be altered in ways nearly imperceptible to the human ear, but sufficient to bypass state-of-the-art  detectors?}

Unlike previous research that focused on natural perturbations~\cite{muller2022does,xie2024codecfake}, we consider a malicious attacker who deliberately optimizes the perturbation to evade detection.
We examine this scenario under various levels of access to the detection systems, from having full knowledge (white-box), to partial knowledge (black-box), to no knowledge (agnostic).

\subsection{Experiment Setup}

We train four SOTA open-source SSDs from scratch: AASIST~\cite{aasist}, AASIST-L~\cite{aasist}, RawNet2~\cite{rawnet2} and RawGATST~\cite{rawnetgatst} on ASVSpoof2019-LA train split~\cite{asvspoof2019}.
Their equal error rates (EERs) without attacks on ASVSpoof2019-LA test split  are reported in Table~\ref{tab:eer}, and closely match the reported numbers in their original papers.

\begin{table}[t]
\caption{Baseline EERs of SSDs on the ASVSpoof2019-LA test split without attacks.}
\label{tab:eer}
\begin{center}
\begin{tabular}{c|c|c|c}
AASIST & AASIST-L & RawNet2 & RawGATST \\
\hline
0.83\% & 0.99\% & 4.88\% & 3.29\% \\
\hline
\end{tabular}
\end{center}
\end{table}

We launch attacks on three synthetic datasets: ASVSpoof2019-LA test split~\cite{asvspoof2019}, WaveFake~\cite{frank2021wavefake} and In-the-wild~\cite{muller2022does}.
In consideration of compute resources, we randomly sub-sample 100 examples from each dataset for the attacks.

We use the attack success rate (\emph{i.e.} the ratio of attacked examples bypassing the target detector) to measure the effectiveness of the attacks.
To ensure the attack does not degrade audio quality, we use both VisQOL~\cite{visqol} and human ratings to confirm that the attacked audio still sounds similar to the original synthetic audio, which we refer to as ``stealthiness''.

\subsection{White-box Attack}

\begin{boxH}
\noindent\textbf{Takeaway:}
\begin{itemize}[leftmargin=15pt]
\item Existing open-source SSDs are extremely vulnerable to white-box attacks.
\item Deepfake speech from TTS not seen during training is more likely to bypass SSDs.
\item White-box attacks can be highly effective and stealthy simultaneously.
\end{itemize}
\end{boxH}

\begin{algorithm}
\caption{White-box Attacks: PGD and I-FGSM}
\label{alg:white_box}
\begin{algorithmic}[1]
\Require Waveform audio $s \in \mathbb{R}^T$, SSD model $f$, perturbation step size $\alpha$, maximum number of iterations $T$, real audio class $\mathcal{R}$, attack $A\in$ \{`PGD', `I-FGSM'\}, $\ell_{\infty}$-norm constraint $\epsilon$
\Ensure Adversarial perturbation $\delta \in \mathbb{R}^T$
\State $\delta \gets 0$ \Comment{initialization}
\For{$t \gets 1$ to $T$}
\If{$A$ == `PGD'}
\State  $\delta \gets \delta - \alpha \cdot \nabla_{\delta} l_{CE}(f(s+\delta),\mathcal{R}) $
\Comment{gradient descent with cross entropy loss $l_{CE}$}
\EndIf
\If{$A$ == `I-FGSM'}
\State  $\delta \gets \delta - \alpha \cdot \text{sign} (\nabla_{\delta} l_{CE}(f(s+\delta),\mathcal{R})) $
\Comment{FGSM with cross entropy loss $l_{CE}$}
\EndIf
\State $\delta \gets \text{clip} (\delta, -\epsilon, \epsilon) $
\Comment{projection}
\If{$f(s+\delta) == \mathcal{R}$} \Comment{attack success}
\State \textbf{Break}
\EndIf
\EndFor
\State \textbf{return} $\delta$
\end{algorithmic}
\end{algorithm}

\input{fig/whitebox_pgd_lr}
\input{fig/whitebox_pgd_L_infty_norm}
\input{fig/whitebox_pgd_iteration}

\begin{table}
\caption{Human ratings of speaker similarity between the original and PGD attacked audio.}
\label{tab:pgd_human}
\centering
\begin{tabular}{l|c|c|c}
 & ASVspoof & WaveFake & In-the-wild \\ \hline
AASIST & 0.970 $\pm$ 0.063 & 0.971 $\pm$ 0.046 & 0.985 $\pm$ 0.030 \\ \hline
AASIST-L & 0.979 $\pm$ 0.037 & 0.979 $\pm$ 0.045 & 0.975 $\pm$ 0.036 \\ \hline
RawNet2 & 0.971 $\pm$ 0.077 & 1.000 $\pm$ 0.000 & 0.967 $\pm$ 0.063 \\ \hline
RawGATST & 0.997 $\pm$ 0.008 & 0.986 $\pm$ 0.030 & 0.997 $\pm$ 0.008 \\ \hline
\end{tabular}
\end{table}

We first study white-box attack, where the adversary has full access to the model.
We choose two white-box attacks: Projected Gradient Descent (PGD)~\cite{mkadry2017towards} and I-FGSM~\cite{kurakin2018adversarial}.
Algorithm~\ref{alg:white_box} shows details of PGD and I-FGSM.

\paragraph{Projected Gradient Descent:} PGD crafts adversarial examples by iteratively taking small steps in the direction that maximizes the model's error, while projecting the perturbed example back within a certain boundary around the original input to maintain a balance between attack success rate and stealthiness. 

PGD has three major hyper-parameters: perturbation step size, $\ell_\infty$-norm constraint, and the number of iterations.
We conduct hyper-parameter search and summarize the results in Figure~\ref{fig:whitebox_pgd_lr},~\ref{fig:whitebox_pgd_norm}, and \ref{fig:white_box_pgd_iteration}.

In Figure~\ref{fig:whitebox_pgd_lr}, we can tell that on WaveFake and In-the-wild, the attack success rate is almost always 100\% while on ASVSpoof2019-LA test the attack success rate hovers between 60\% and 100\% depending on the learning rate used.
This reflects the fact that the detectors are more robust on test data generated by the same TTS systems as the training data (\emph{i.e.} in-domain data), but are still vulnerable under white-box attacks with a few steps of hyper-parameter search.
On the other hand, VisQOL scores keep decreasing as the perturbation step size grows.
Usually, VisQOL score above 3.0 is considered reasonable quality.
Thus, there exists a sweet spot of perturbation step size striking balance between attack effectiveness and stealthiness.

In Figure~\ref{fig:whitebox_pgd_norm}, the observation of SSDs being more robust on ASVSpoof2019-LA test holds true.
However, we observe that the VisQOL scores are pretty consistent despite the changing $\ell_\infty$-norm constraint, which says that audio quality is insensitive to $\ell_\infty$-norm constraint within a certain range.

Figure~\ref{fig:white_box_pgd_iteration} shows that white-box attacks are efficient, reaching maximum attack success rates and stable VisQOL scores after just 50 iterations.

We also collect human ratings on whether the PGD-attacked audio with the best hyper-parameter combination sounds like the original synthetic audio, and the results are summarized in Table~\ref{tab:pgd_human}.
We can see that most human raters think the two audio sound like the same person, underscoring the potential threat of using the attacked audio for impersonation.

\paragraph{Iterative Fast Gradient Sign Method:}
I-FGSM only differs from PGD in that it only uses the sign of the gradient to perturb the input audio. It shares the same set of hyper-parameters as PGD, for which the grid search results are summarized in Figure~\ref{fig:whitebox_fgsm_lr},~\ref{fig:whitebox_fgsm_norm}, and~\ref{fig:white_box_fgsm_iteration} and human ratings are summarized in Table~\ref{tab:fgms_human}, and the findings are similar to PGD.

\input{fig/whitebox_fgsm_lr}
\input{fig/whitebox_fgsm_L_infty_norm}
\input{fig/whitebox_fgsm_iteration}

\begin{table}[b]
\vspace{-5pt}
\caption{Human ratings of speaker similarity between the original and I-FGSM attacked audio.}
\label{tab:fgms_human}
\centering
\begin{tabular}{l|c|c|c}
 & ASVspoof & WaveFake & In-the-wild \\ \hline
AASIST & 0.984 $\pm$ 0.020 & 0.960 $\pm$ 0.052 & 0.985 $\pm$ 0.024 \\ \hline
AASIST-L & 0.987 $\pm$ 0.022 & 0.986 $\pm$ 0.023 & 0.967 $\pm$ 0.054 \\ \hline
RawNet2 & 0.980 $\pm$ 0.040 & 1.000 $\pm$ 0.000 & 0.991 $\pm$ 0.012 \\ \hline
RawGATST & 0.989 $\pm$ 0.024 & 0.858 $\pm$ 0.141 & 0.985 $\pm$ 0.024 \\ \hline
\end{tabular}
\vspace{-5pt}
\end{table}

\subsection{Black-box Attack}

\begin{boxH}
\noindent\textbf{Takeaway:}
\begin{itemize}[leftmargin=15pt]
\item Existing open-source SSDs are still vulnerable to black-box attacks.
\item Black-box attacks can be effective and stealthy simultaneously.
\end{itemize}
\end{boxH}

\input{fig/SimBA_pert_batch_size}
\input{fig/SimBA_epsilon}
\begin{figure}[!t]

\centering
\begin{subfigure}[b]{0.99\textwidth}
\centering
    \begin{adjustbox}{width=0.3\linewidth}
    \begin{tikzpicture}
    \begin{axis}[
        title={ASVSpoof2019-LA},
        title style={font=\fontsize{14}{12}\selectfont},
        grid=both,
        cycle list/Dark2,
        mark list fill={.!75!white},
        cycle multiindex* list={
            Dark2\nextlist
            my marks\nextlist
            dashed\nextlist
            very thick\nextlist
        },
        every axis plot/.append style={
            errorBars,
            mark size=4pt,
            line width=5pt,
        },
        xlabel={\#Queries},
        ylabel={Attack Success Rate},
        ticklabel style={font=\fontsize{14}{12}\selectfont},
        label style={font=\fontsize{14}{12}\selectfont},
        legend entries={AASIST, AASIST-L, RawNet2, RawGATST},
        legend style={at={(0.01,0.01)}, anchor=south west, fill=white, fill opacity=0.6, text opacity=1, font=\fontsize{14}{12}\selectfont},
        xtick={1000,2500,5000,7500},
        xticklabels={1000,2500,5000,7500},
        % xmode=log,
        ymin=0,ymax=1.05,
    ]

    \addplot table[col sep=comma, x=query, y=asr-avg, y error=asr-std] {data/SimBA_impact_query/aasist.txt};
    \addplot table[col sep=comma, x=query, y=asr-avg, y error=asr-std] {data/SimBA_impact_query/aasist_l.txt};
    \addplot table[col sep=comma, x=query, y=asr-avg, y error=asr-std] {data/SimBA_impact_query/rawnet2.txt};
    \addplot table[col sep=comma, x=query, y=asr-avg, y error=asr-std] {data/SimBA_impact_query/rawgatst.txt};

    \end{axis}
    \end{tikzpicture}
    \end{adjustbox}
    \begin{adjustbox}{width=0.3\linewidth}
    \begin{tikzpicture}
    \begin{axis}[
        title={WaveFake},
        title style={font=\fontsize{14}{12}\selectfont},
        grid=both,
        cycle list/Dark2,
        mark list fill={.!75!white},
        cycle multiindex* list={
            Dark2\nextlist
            my marks\nextlist
            dashed\nextlist
            very thick\nextlist
        },
        every axis plot/.append style={
            errorBars,
            mark size=4pt,
            line width=5pt,
        },
        xlabel={\#Queries},
        ylabel={Attack Success Rate},
        ticklabel style={font=\fontsize{14}{12}\selectfont},
        label style={font=\fontsize{14}{12}\selectfont},
        legend entries={AASIST, AASIST-L, RawNet2, RawGATST},
        legend style={at={(0.01,0.01)}, anchor=south west, fill=white, fill opacity=0.6, text opacity=1, font=\fontsize{14}{12}\selectfont},
        xtick={1000,2500,5000,7500},
        xticklabels={1000,2500,5000,7500},
        % xmode=log,
        ymin=0,ymax=1.05,
    ]

    \addplot table[col sep=comma, x=query, y=asr-avg, y error=asr-std] {data/SimBA_impact_query_wavefake/aasist.txt};
    \addplot table[col sep=comma, x=query, y=asr-avg, y error=asr-std] {data/SimBA_impact_query_wavefake/aasist_l.txt};
    \addplot table[col sep=comma, x=query, y=asr-avg, y error=asr-std] {data/SimBA_impact_query_wavefake/rawnet2.txt};
    \addplot table[col sep=comma, x=query, y=asr-avg, y error=asr-std] {data/SimBA_impact_query_wavefake/rawgatst.txt};

    \end{axis}
    \end{tikzpicture}
    \end{adjustbox}
    \begin{adjustbox}{width=0.3\linewidth}
    \begin{tikzpicture}
    \begin{axis}[
        title={In-the-Wild},
        title style={font=\fontsize{14}{12}\selectfont},
        grid=both,
        cycle list/Dark2,
        mark list fill={.!75!white},
        cycle multiindex* list={
            Dark2\nextlist
            my marks\nextlist
            dashed\nextlist
            very thick\nextlist
        },
        every axis plot/.append style={
            errorBars,
            mark size=4pt,
            line width=5pt,
        },
        xlabel={\#Queries},
        ylabel={Attack Success Rate},
        ticklabel style={font=\fontsize{14}{12}\selectfont},
        label style={font=\fontsize{14}{12}\selectfont},
        legend entries={AASIST, AASIST-L, RawNet2, RawGATST},
        legend style={at={(0.01,0.01)}, anchor=south west, fill=white, fill opacity=0.6, text opacity=1, font=\fontsize{14}{12}\selectfont},
        xtick={1000,2500,5000,7500},
        xticklabels={1000,2500,5000,7500},
        % xmode=log,
        ymin=0,ymax=1.05,
    ]

    \addplot table[col sep=comma, x=query, y=asr-avg, y error=asr-std] {data/SimBA_impact_query_wild/aasist.txt};
    \addplot table[col sep=comma, x=query, y=asr-avg, y error=asr-std] {data/SimBA_impact_query_wild/aasist_l.txt};
    \addplot table[col sep=comma, x=query, y=asr-avg, y error=asr-std] {data/SimBA_impact_query_wild/rawnet2.txt};
    \addplot table[col sep=comma, x=query, y=asr-avg, y error=asr-std] {data/SimBA_impact_query_wild/rawgatst.txt};

    \end{axis}
    \end{tikzpicture}
    \end{adjustbox}

    \caption{Attack success rate vs. SimBA \#queries.}
    \label{fig:simba_asr_query}
\end{subfigure}

\centering
\begin{subfigure}[b]{0.99\textwidth}
\centering
    \begin{adjustbox}{width=0.3\linewidth}
    \begin{tikzpicture}
    \begin{axis}[
        title={ASVSpoof2019-LA},
        title style={font=\fontsize{14}{12}\selectfont},
        grid=both,
        cycle list/Dark2,
        mark list fill={.!75!white},
        cycle multiindex* list={
            Dark2\nextlist
            my marks\nextlist
            dashed\nextlist
            very thick\nextlist
        },
        every axis plot/.append style={
            errorBars,
            mark size=4pt,
            line width=5pt,
        },
        xlabel={\#Queries},
        ylabel={ViSQOL},
        ticklabel style={font=\fontsize{14}{12}\selectfont},
        label style={font=\fontsize{14}{12}\selectfont},
        legend entries={AASIST, AASIST-L, RawNet2, RawGATST},
        legend style={at={(0.01,0.01)}, anchor=south west, fill=white, fill opacity=0.6, text opacity=1, font=\fontsize{14}{12}\selectfont},
        xtick={1000,2500,5000,7500},
        xticklabels={1000,2500,5000,7500},
        % xmode=log,
        ymin=0, ymax=5.25,
    ]

    \addplot table[col sep=comma, x=query, y=visqol-avg, y error=visqol-std] {data/SimBA_impact_query/aasist.txt};
    \addplot table[col sep=comma, x=query, y=visqol-avg, y error=visqol-std] {data/SimBA_impact_query/aasist_l.txt};
    \addplot table[col sep=comma, x=query, y=visqol-avg, y error=visqol-std] {data/SimBA_impact_query/rawnet2.txt};
    \addplot table[col sep=comma, x=query, y=visqol-avg, y error=visqol-std] {data/SimBA_impact_query/rawgatst.txt};

    \end{axis}
    \end{tikzpicture}
    \end{adjustbox}
    \begin{adjustbox}{width=0.3\linewidth}
    \begin{tikzpicture}
    \begin{axis}[
        title={WaveFake},
        title style={font=\fontsize{14}{12}\selectfont},
        grid=both,
        cycle list/Dark2,
        mark list fill={.!75!white},
        cycle multiindex* list={
            Dark2\nextlist
            my marks\nextlist
            dashed\nextlist
            very thick\nextlist
        },
        every axis plot/.append style={
            errorBars,
            mark size=4pt,
            line width=5pt,
        },
        xlabel={\#Queries},
        ylabel={ViSQOL},
        ticklabel style={font=\fontsize{14}{12}\selectfont},
        label style={font=\fontsize{14}{12}\selectfont},
        legend entries={AASIST, AASIST-L, RawNet2, RawGATST},
        legend style={at={(0.01,0.01)}, anchor=south west, fill=white, fill opacity=0.6, text opacity=1, font=\fontsize{14}{12}\selectfont},
        xtick={1000,2500,5000,7500},
        xticklabels={1000,2500,5000,7500},
        % xmode=log,
        ymin=0, ymax=5.25,
    ]

    \addplot table[col sep=comma, x=query, y=visqol-avg, y error=visqol-std] {data/SimBA_impact_query_wavefake/aasist.txt};
    \addplot table[col sep=comma, x=query, y=visqol-avg, y error=visqol-std] {data/SimBA_impact_query_wavefake/aasist_l.txt};
    \addplot table[col sep=comma, x=query, y=visqol-avg, y error=visqol-std] {data/SimBA_impact_query_wavefake/rawnet2.txt};
    \addplot table[col sep=comma, x=query, y=visqol-avg, y error=visqol-std] {data/SimBA_impact_query_wavefake/rawgatst.txt};

    \end{axis}
    \end{tikzpicture}
    \end{adjustbox}
    \begin{adjustbox}{width=0.3\linewidth}
    \begin{tikzpicture}
    \begin{axis}[
        title={In-the-Wild},
        title style={font=\fontsize{14}{12}\selectfont},
        grid=both,
        cycle list/Dark2,
        mark list fill={.!75!white},
        cycle multiindex* list={
            Dark2\nextlist
            my marks\nextlist
            dashed\nextlist
            very thick\nextlist
        },
        every axis plot/.append style={
            errorBars,
            mark size=4pt,
            line width=5pt,
        },
        xlabel={\#Queries},
        ylabel={ViSQOL},
        ticklabel style={font=\fontsize{14}{12}\selectfont},
        label style={font=\fontsize{14}{12}\selectfont},
        legend entries={AASIST, AASIST-L, RawNet2, RawGATST},
        legend style={at={(0.01,0.01)}, anchor=south west, fill=white, fill opacity=0.6, text opacity=1, font=\fontsize{14}{12}\selectfont},
        xtick={1000,2500,5000,7500},
        xticklabels={1000,2500,5000,7500},
        % xmode=log,
        ymin=0, ymax=5.25,
    ]

    \addplot table[col sep=comma, x=query, y=visqol-avg, y error=visqol-std] {data/SimBA_impact_query_wild/aasist.txt};
    \addplot table[col sep=comma, x=query, y=visqol-avg, y error=visqol-std] {data/SimBA_impact_query_wild/aasist_l.txt};
    \addplot table[col sep=comma, x=query, y=visqol-avg, y error=visqol-std] {data/SimBA_impact_query_wild/rawnet2.txt};
    \addplot table[col sep=comma, x=query, y=visqol-avg, y error=visqol-std] {data/SimBA_impact_query_wild/rawgatst.txt};

    \end{axis}
    \end{tikzpicture}
    \end{adjustbox}
\caption{ViSQOL score vs. SimBA \#queries.}
\label{fig:simba_visqol_query}
\end{subfigure}

\caption{Attack success rate and ViSQOL vs. SimBA \#queries across datasets.}
\label{fig:simba_query}
\end{figure}
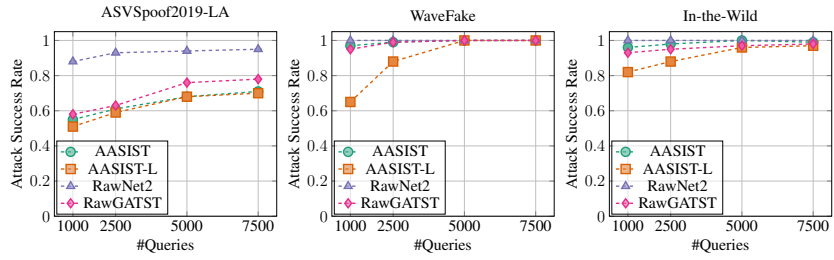
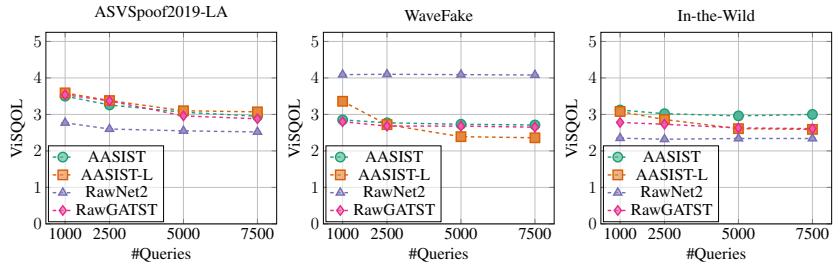

\begin{table}
\caption{Human ratings of speaker similarity between the original and simBA attacked audio.}
\label{tab:simba_human}
\centering
\begin{tabular}{l|c|c|c}
 & ASVspoof & WaveFake & In-the-wild \\ \hline
AASIST & 0.984 $\pm$ 0.020 & 0.960 $\pm$ 0.052 & 0.985 $\pm$ 0.024 \\ \hline
AASIST-L & 0.987 $\pm$ 0.022 & 0.986 $\pm$ 0.023 & 0.967 $\pm$ 0.054 \\ \hline
RawNet2 & 0.980 $\pm$ 0.040 & 1.000 $\pm$ 0.000 & 0.991 $\pm$ 0.012 \\ \hline
RawGATST & 0.989 $\pm$ 0.024 & 0.858 $\pm$ 0.141 & 0.985 $\pm$ 0.024 \\ \hline
\end{tabular}
\end{table}

For black-box attack, we choose the Simple Black Box Attack (SimBA)~\cite{simba}.
SimBA perturbs the input audio randomly and observes whether the prediction confidence score for ``fake'' class decreases or increases.
If the confidence score decreases, SimBA will keep the perturbation.
Otherwise, SimBA will try adding perturbation in the opposite direction and decide whether to keep or discard the perturbation just as above.
SimBA iteratively perturbs the input until the SSD is successfully bypassed or the budget of queries/iterations is used up.
Algorithm~\ref{alg:black_box} shows the details of SimBA. 

SimBA has three hyper-parameters: perturbation batch size, perturbation step size, and the number of queries.
Perturbation batch size decides how many timesteps are perturbed in each query, while perturbation step size decides which long one perturbation step on one timestep can be.
The hyper-parameter search results are summarized in Figure~\ref{fig:simba_pert_batch_size},~\ref{fig:simba_pert_step_size}, and~\ref{fig:simba_query}.

In Figure~\ref{fig:simba_pert_batch_size}, we observe that on ASVSpoof2019 test, RawNet2, the least capable SSD model is still broken almost 100\% but all the other 3 models are only broken 60\% of all the tested examples.
This draws a positive correlation between model capability and robustness.
On WaveFake and In-the-Wild, all SSDs are broken more than 90\% of the time, which confirms the previous observation that current SSD models are brittle when facing synthetic audio from TTS systems never seen during training.

Also in Figure~\ref{fig:simba_pert_batch_size},~\ref{fig:simba_pert_step_size}, and~\ref{fig:simba_query}, we observe that ASSIST-L is the most robust model consistently, which is surprising because it's the smallest model within the 4 (See \cite{aasist} for the size of these models.).
This observation aligns with the principle of Occam's razor, which suggests that simpler models often generalize better.
A potential explanation could lie in the raggedness of the decision boundaries.
Larger models, with their increased complexity, might create more intricate and potentially overfit decision boundaries.
In contrast, ASSIST-L, being smaller, may form smoother decision boundaries, leading to better generalization and robustness against perturbations.

Human ratings of audio similarity is summarized in Table~\ref{tab:simba_human}.
Again the attacked audio sound highly similar to the original ones to human ears.

\begin{algorithm}[t]
\caption{Black-box Attack: SimBA}
\label{alg:black_box}
\begin{algorithmic}[1]
\Require Waveform audio $s \in \mathbb{R}^T$, SSD model $f$, perturbation step size $\alpha$, perturbation batch size $q$, maximum number of queries $Q$, real audio class $\mathcal{R}$, $\ell_{\infty}$-norm constraint $\epsilon$
\Ensure Adversarial perturbation $\delta \in \mathbb{R}^T$
\State $\delta \gets 0, t \gets 0$ \Comment{initialization}
\State $p \gets f(s, \mathcal{R}), t \gets t + 1$ \Comment{initializing highest probability to predict real audio class}
\While{$t < T$}
\If{$f(s+\delta)$ == $\mathcal{R}$} \Comment{attack success}
\State \textbf{Break}
\EndIf
\State $r \in \mathbb{R}^T$ and $r \gets 0 $    
\State Randomly choose $q$ dimensions from $r$ without replacement
\State Randomly add $\alpha$ or $-\alpha$ to the chosen $q$ dimensions in $r$
\State $t \gets t + 1$  \Comment{one more query for $f$ below}
\If{$f(s+\delta+r,\mathcal{R}) > p $}
\State $p \gets f(s+\delta+r,\mathcal{R})$  \Comment{update highest probability to predict real audio class}
\State \textbf{Continue}
\Else
\State $t \gets t + 1$  \Comment{one more query for $f$ below}
\If{$f(s+\delta-r,\mathcal{R}) > p $}
\State $p \gets f(s+\delta-r,\mathcal{R})$  \Comment{update highest probability to predict real audio class}
\State \textbf{Continue}
\EndIf
\EndIf
\EndWhile
\State \textbf{return} $\delta$
\end{algorithmic}
\end{algorithm}

\subsection{Agnostic Attack: Transferability of Above Attacks}

\begin{boxH}
\noindent\textbf{Takeaway:}
\begin{itemize}[leftmargin=15pt]
\item For both white-box attacks and black-box attacks, transferability depends on the target model's capability on the target audio.
\item Black-box attacks are more transferrable on in-domain test data than out-of-domain data.
\item Transferrability of different white-box attacks are alike.
\end{itemize}
\end{boxH}

The above attacks all assume different levels of access to the SSD model which might not be accessible in practice.
As a result, we want to understand whether the above attacks are transferrable: Can a successfully attacked example on one model transfer to a different model?
If this is true, then the adversary can craft a proxy model themselves, attack it, and expect it to bypass the real SSD as well.

\begin{figure}[!t]
    \centering

    \subfloat[PGD on ASVSpoof2019-LA]{
        \includegraphics[width=0.33\linewidth]{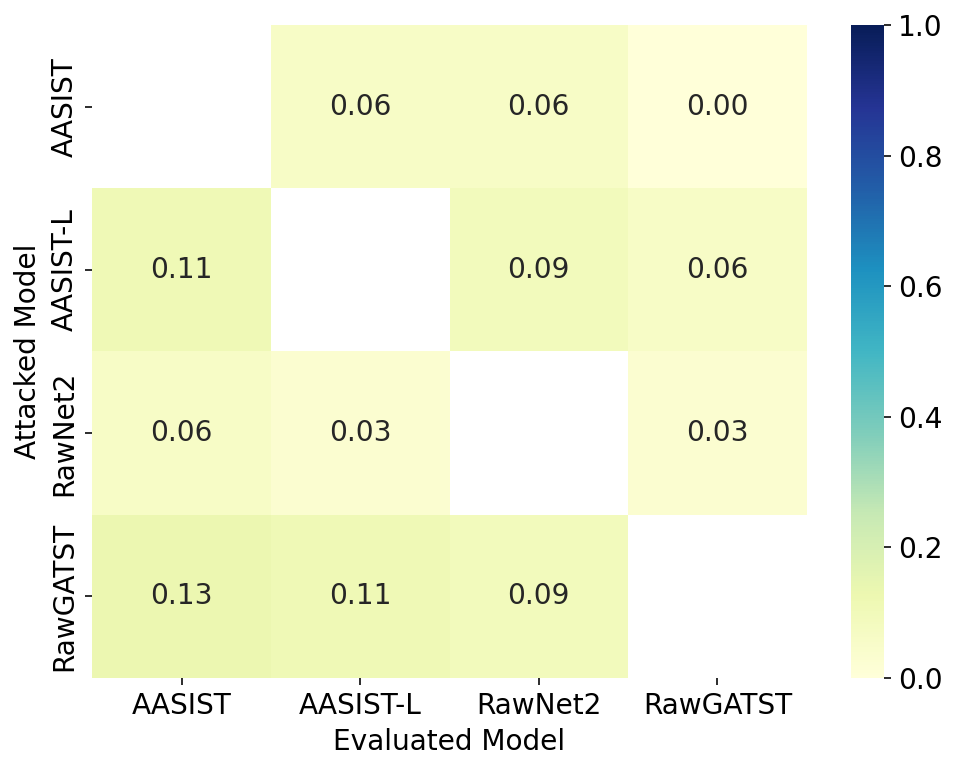}
        \label{fig:transferability_asvspoof1}
    }
    \subfloat[PGD on WaveFake]{
        \includegraphics[width=0.33\linewidth]{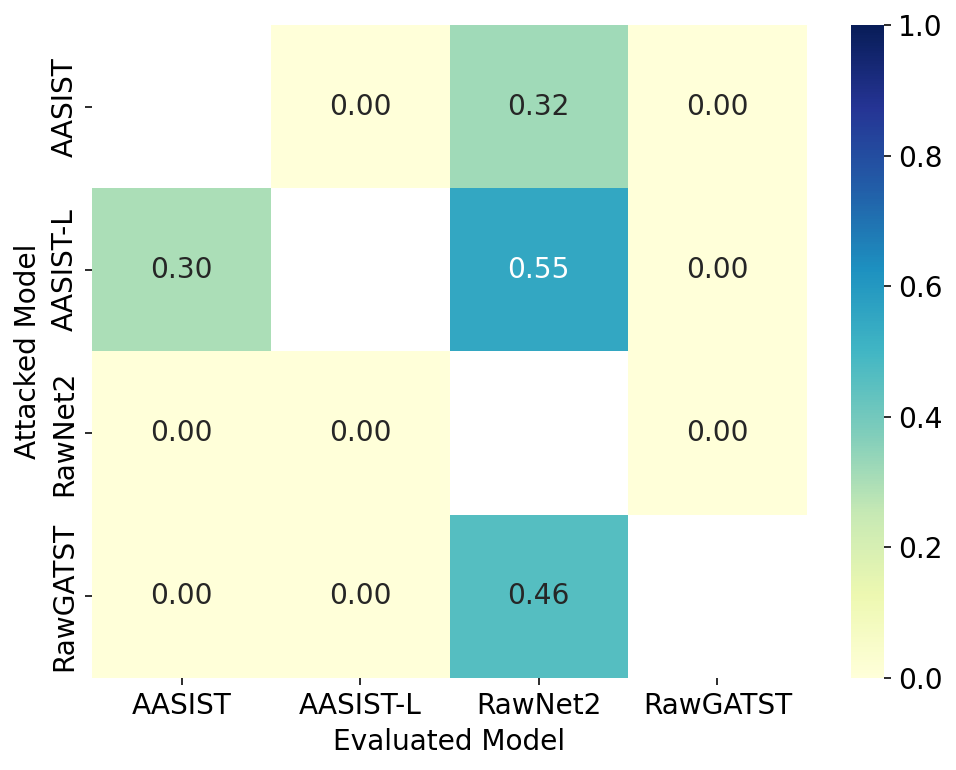}
        \label{fig:transferability_wavefake}
    }
    \subfloat[PGD on In-the-Wild]{
        \includegraphics[width=0.33\linewidth]{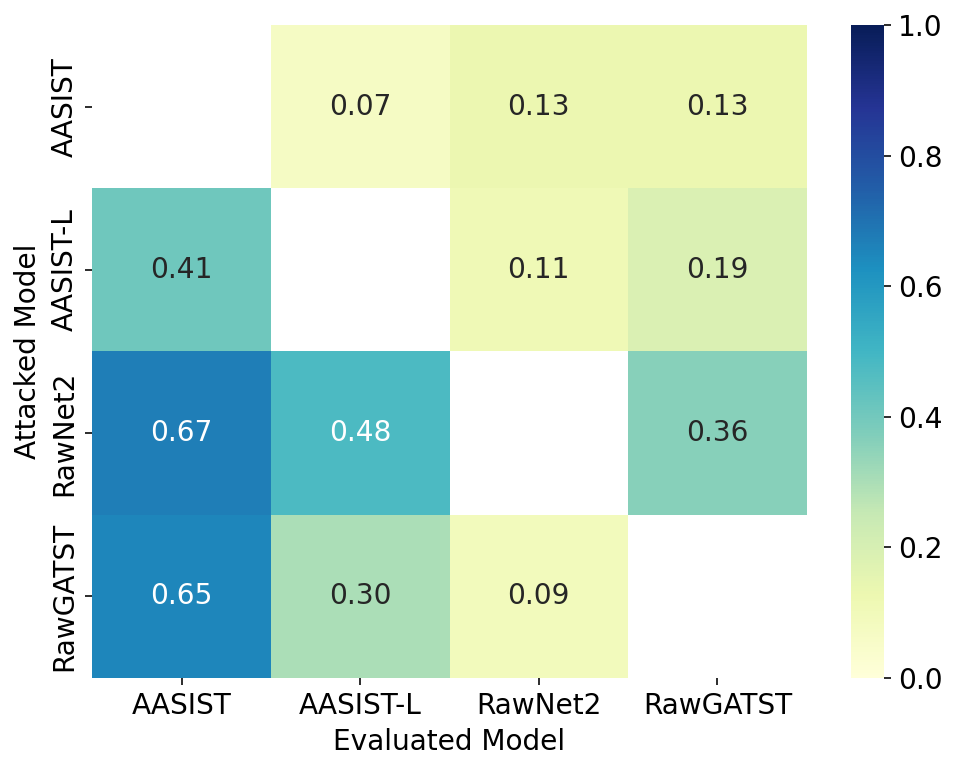}
        \label{fig:transferability_in_the_wild}
    }

    \subfloat[I-FGSM on ASVSpoof2019-LA]{
        \includegraphics[width=0.33\linewidth]{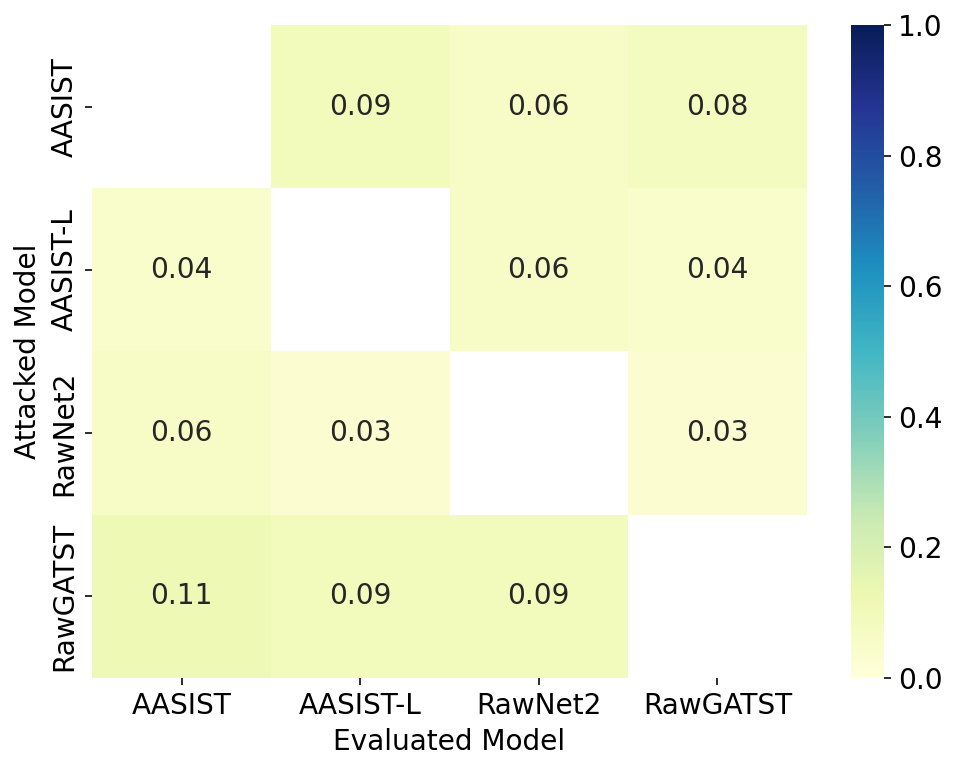}
        \label{fig:transferability_asvspoof1}
    }
    \subfloat[I-FGSM on WaveFake]{
        \includegraphics[width=0.33\linewidth]{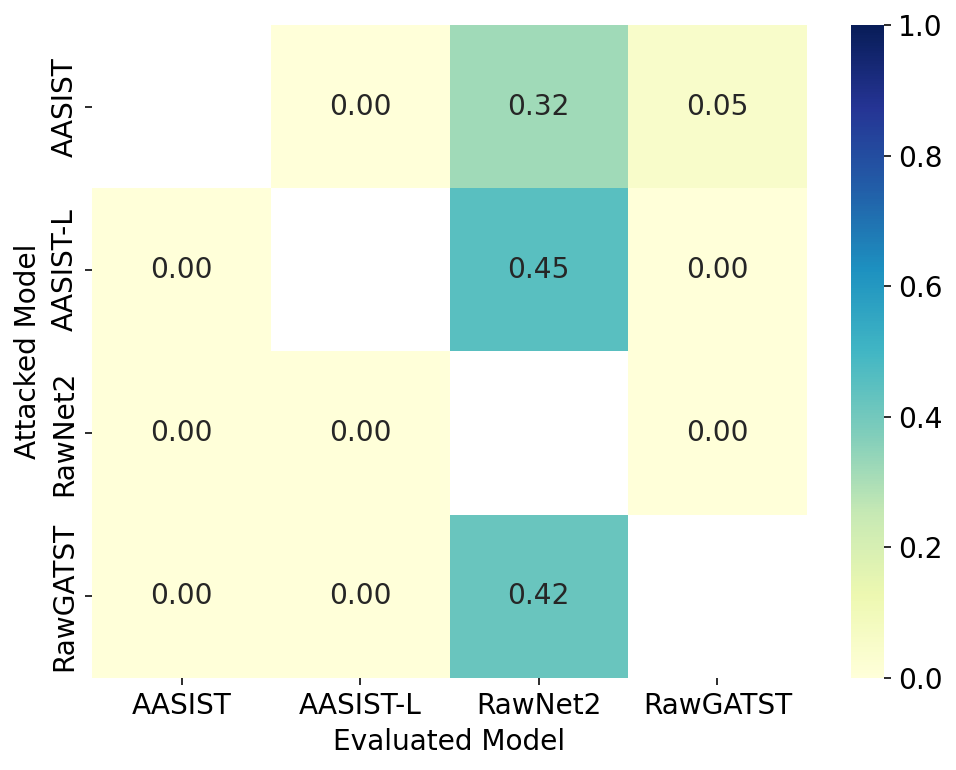}
        \label{fig:transferability_wavefake}
    }
    \subfloat[I-FGSM on In-the-Wild]{
        \includegraphics[width=0.33\linewidth]{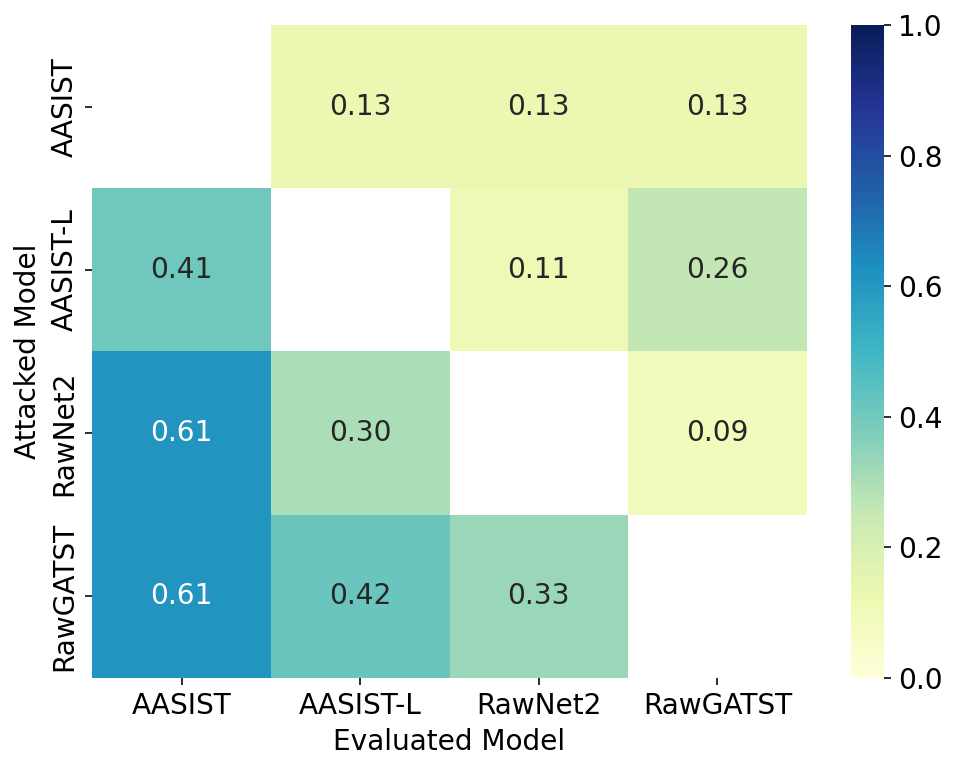}
        \label{fig:transferability_in_the_wild}
    }

    \subfloat[SimBA on ASVSpoof2019-LA]{
        \includegraphics[width=0.33\linewidth]{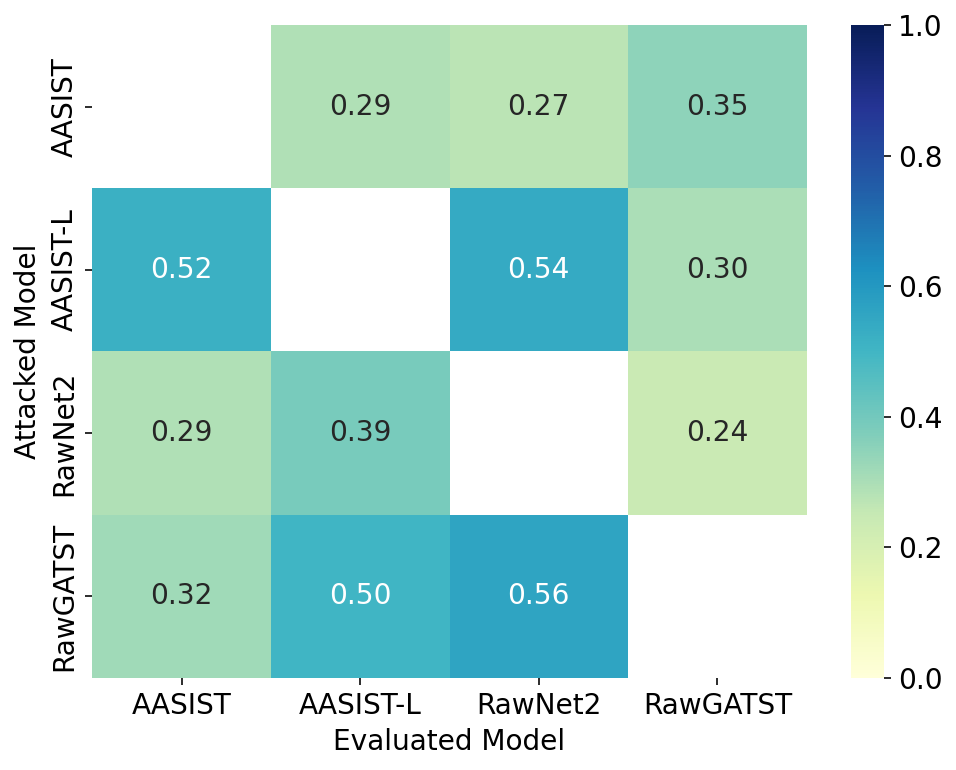}
        \label{fig:transferability_asvspoof1}
    }
    \subfloat[SimBA on WaveFake]{
        \includegraphics[width=0.33\linewidth]{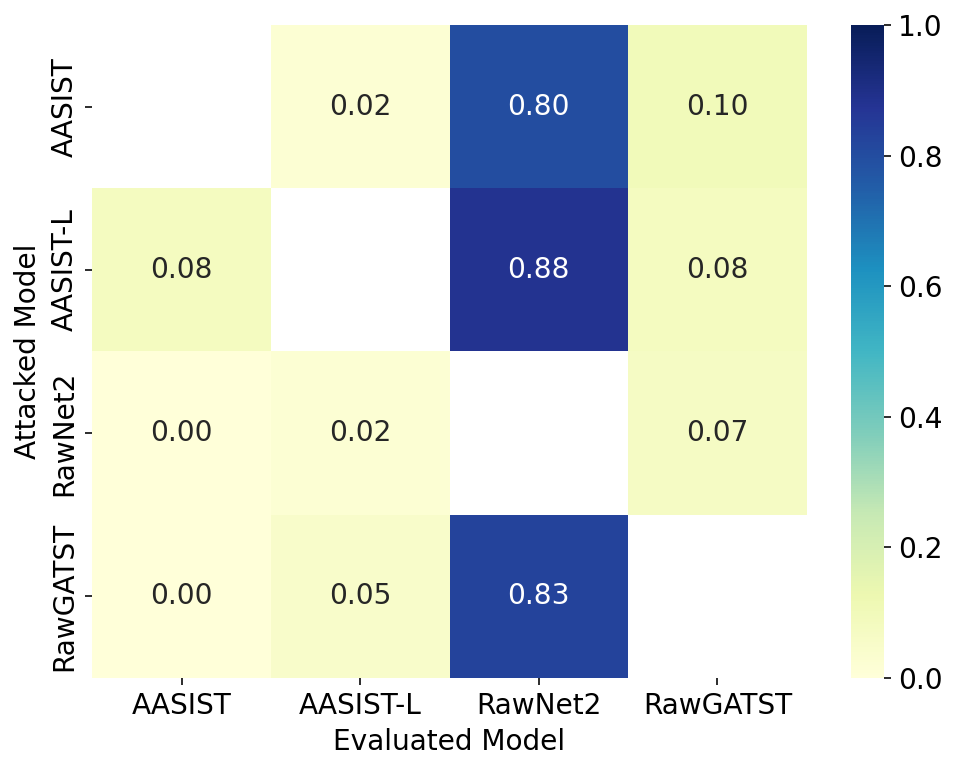}
        \label{fig:transferability_wavefake}
    }
    \subfloat[SimBA on In-the-Wild]{
        \includegraphics[width=0.33\linewidth]{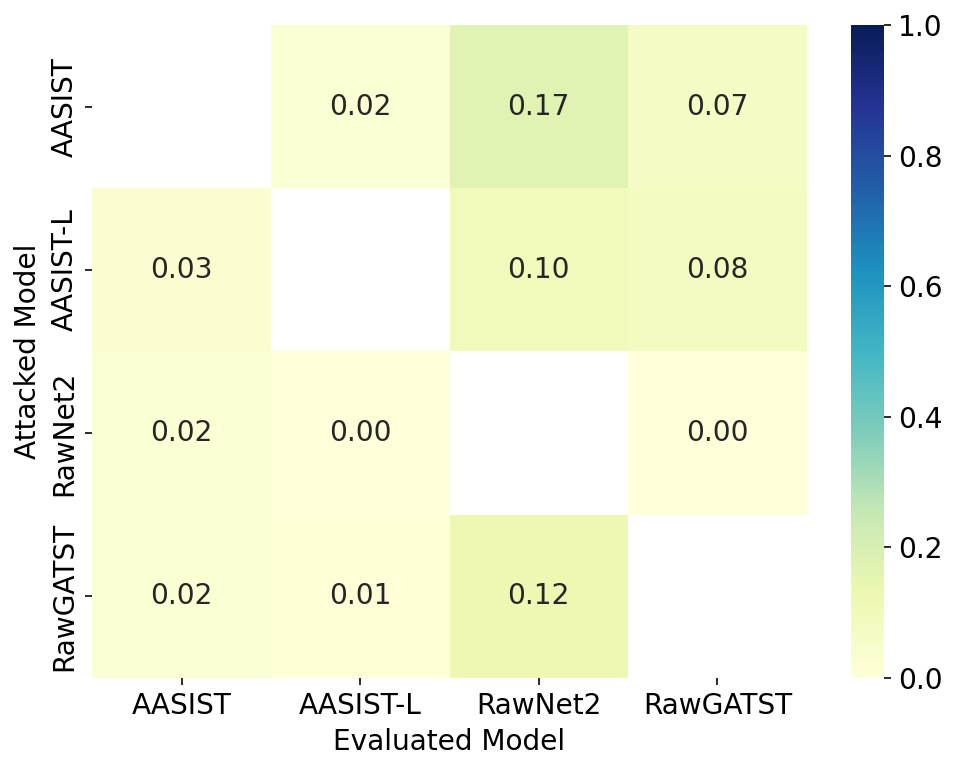}
        \label{fig:transferability_in_the_wild}
    }
    \caption{Transferability of attacks on different datasets.}
    \label{fig:transferability_combined}
\end{figure}

The results are summarized in Figure~\ref{fig:transferability_combined}.
First, we find that on out-of-domain data, some SSDs are extremely vulnerable.
For example, on WaveFake, RawNet2 is extremely vulnerable under all attacks; on In-the-wild, ASSIST and AASIST-L are more vulnerable than the other two models.
Second, we find that on in-domain data, black-box attacks are much more transferrable than white-box attacks.
This is because 1) black-box attacks tend to add larger perturbation than white-box attacks; 2) the SSDs' decision borders are alike for in-domain data.
Thirdly, we also observe high similarity between the transferrability heatmap between PGD and I-FGSM, which might be due to different white-box attacks taking gradient paths in similar directions despite small differences.

\section{Conclusion}
\label{sec:conclusion}

This work presents the first systematic exploration of the robustness of state-of-the-art SSDs against adversarial attacks.
Our findings reveal critical implications for SSD deployment and future research.

Firstly, we demonstrate a clear correlation between system accessibility and vulnerability.  Open-access SSDs, and even those with oracle access, are  highly susceptible to attacks. This underscores the critical need to restrict public access to SSD models and internal workings. While complete prevention of information leakage may be challenging, measures such as rate limiting can effectively mitigate the threat of black-box attacks.

Secondly, our research exposes the vulnerability of current open-source SSDs to domain shifts, including active perturbations and, notably, synthetic data generated from unseen TTS systems. This highlights two key recommendations:

\begin{itemize}
\item Comprehensive and Continuously Updated Training Data: SSD training datasets should encompass a diverse range of TTS systems and be regularly updated to incorporate new systems, ensuring broad coverage against evolving spoofing techniques.
\item Data Composition Confidentiality: Maintaining the confidentiality of training data composition is crucial to prevent attackers from exploiting this knowledge for targeted attacks, as evidenced by the significant advantage gained from TTS system shifts even in agnostic attacks.
\end{itemize}

Our findings also draw a parallel between the interplay of SSD and TTS development and an offline GAN, suggesting a future where TTS systems may achieve near-perfect human mimicry. This emphasizes the crucial need for complementary audio authenticity techniques, such as robust watermarking~\cite{liu2024audiomarkbench}, to bolster the security of SSD systems.

In conclusion, this study serves as a critical analysis of the current state of open-source SSD robustness. By exposing key vulnerabilities and providing actionable recommendations, we aim to guide future research and development efforts towards building more secure and resilient SSD systems in an ever-evolving landscape of speech synthesis and spoofing technologies.

% Hongbin spearheaded the technical development, responsible for the majority of the codebase and the initial conception of the speech-simBA. Joseph played a critical role in the implementation process and led the human evaluation efforts. Arun's expertise was central to the design of speech-simBA method, and his insights proved valuable throughout the project lifecycle. Athula provided crucial guidance from a practitioner's perspective, shaping the project's direction.  Pedro initiated the project and laid the groundwork for its development.  Lun, in collaboration with Pedro, defined the scope of the internship project, oversaw its execution, developed the initial framework subsequently adapted by Hongbin, and propose to use segment silence as the standard to add perturbation in speech-simBA.


\bibliography{ref}

\begin{thebibliography}{10}

\bibitem{kreuk2022audiogen}
Felix Kreuk, Gabriel Synnaeve, Adam Polyak, Uriel Singer, Alexandre
  D{\'e}fossez, Jade Copet, Devi Parikh, Yaniv Taigman, and Yossi Adi.
\newblock Audiogen: Textually guided audio generation.
\newblock {\em arXiv preprint arXiv:2209.15352}, 2022.

\bibitem{borsos2023audiolm}
Zal{\'a}n Borsos, Rapha{\"e}l Marinier, Damien Vincent, Eugene Kharitonov,
  Olivier Pietquin, Matt Sharifi, Dominik Roblek, Olivier Teboul, David
  Grangier, Marco Tagliasacchi, et~al.
\newblock Audiolm: a language modeling approach to audio generation.
\newblock {\em IEEE/ACM transactions on audio, speech, and language
  processing}, 31:2523--2533, 2023.

\bibitem{leng2023prompttts}
Yichong Leng, Zhifang Guo, Kai Shen, Xu~Tan, Zeqian Ju, Yanqing Liu, Yufei Liu,
  Dongchao Yang, Leying Zhang, Kaitao Song, et~al.
\newblock Prompttts 2: Describing and generating voices with text prompt.
\newblock {\em arXiv preprint arXiv:2309.02285}, 2023.

\bibitem{saeki2023virtuoso}
Takaaki Saeki, Heiga Zen, Zhehuai Chen, Nobuyuki Morioka, Gary Wang, Yu~Zhang,
  Ankur Bapna, Andrew Rosenberg, and Bhuvana Ramabhadran.
\newblock Virtuoso: Massive multilingual speech-text joint semi-supervised
  learning for text-to-speech.
\newblock In {\em ICASSP 2023-2023 IEEE International Conference on Acoustics,
  Speech and Signal Processing (ICASSP)}, pages 1--5. IEEE, 2023.

\bibitem{shen2023naturalspeech}
Kai Shen, Zeqian Ju, Xu~Tan, Yanqing Liu, Yichong Leng, Lei He, Tao Qin, Sheng
  Zhao, and Jiang Bian.
\newblock Naturalspeech 2: Latent diffusion models are natural and zero-shot
  speech and singing synthesizers.
\newblock {\em arXiv preprint arXiv:2304.09116}, 2023.

\bibitem{wang2023valle}
Chengyi Wang, Sanyuan Chen, Yu~Wu, Ziqiang Zhang, Long Zhou, Shujie Liu, Zhuo
  Chen, Yanqing Liu, Huaming Wang, Jinyu Li, et~al.
\newblock Neural codec language models are zero-shot text to speech
  synthesizers.
\newblock {\em arXiv preprint arXiv:2301.02111}, 2023.

\bibitem{chattts}
ChatTTS.
\newblock Chattts - text-to-speech for conversational scenarios.
\newblock https://chattts.com/, 2024.
\newblock Accessed: 07/16/2024.

\bibitem{chen2024valle2}
Sanyuan Chen, Shujie Liu, Long Zhou, Yanqing Liu, Xu~Tan, Jinyu Li, Sheng Zhao,
  Yao Qian, and Furu Wei.
\newblock Vall-e 2: Neural codec language models are human parity zero-shot
  text to speech synthesizers.
\newblock {\em arXiv preprint arXiv:2406.05370}, 2024.

\bibitem{coqui}
xTTS.
\newblock High quality, human-like ai voice generator.
\newblock https://github.com/coqui-ai/TTS, 2024.
\newblock Accessed: 07/16/2024.

\bibitem{elevenlabs}
ElevenLabs.
\newblock High quality, human-like ai voice generator.
\newblock \url{https://elevenlabs.io/text-to-speech}, 2024.
\newblock Accessed: 07/16/2024.

\bibitem{le2024voicebox}
Matthew Le, Apoorv Vyas, Bowen Shi, Brian Karrer, Leda Sari, Rashel Moritz,
  Mary Williamson, Vimal Manohar, Yossi Adi, Jay Mahadeokar, et~al.
\newblock Voicebox: Text-guided multilingual universal speech generation at
  scale.
\newblock {\em Advances in neural information processing systems}, 36, 2024.

\bibitem{lux2024massive}
Florian Lux, Sarina Meyer, Lyonel Behringer, Frank Zalkow, Phat Do, Matt Coler,
  Emanuël A.~P. Habets, and Ngoc~Thang Vu.
\newblock {Meta Learning Text-to-Speech Synthesis in over 7000 Languages}.
\newblock In {\em Interspeech}. ISCA, 2024.

\bibitem{biadsy2024zero}
Fadi Biadsy, Youzheng Chen, Isaac Elias, Kyle Kastner, Gary Wang, Andrew
  Rosenberg, and Bhuvana Ramabhadran.
\newblock Zero-shot cross-lingual voice transfer for tts.
\newblock {\em arXiv preprint arXiv:2409.13910}, 2024.

\bibitem{cooper2020zero}
Erica Cooper, Cheng-I Lai, Yusuke Yasuda, Fuming Fang, Xin Wang, Nanxin Chen,
  and Junichi Yamagishi.
\newblock Zero-shot multi-speaker text-to-speech with state-of-the-art neural
  speaker embeddings.
\newblock In {\em ICASSP 2020-2020 IEEE International Conference on Acoustics,
  Speech and Signal Processing (ICASSP)}, pages 6184--6188. IEEE, 2020.

\bibitem{casanova2022yourtts}
Edresson Casanova, Julian Weber, Christopher~D Shulby, Arnaldo~Candido Junior,
  Eren G{\"o}lge, and Moacir~A Ponti.
\newblock Yourtts: Towards zero-shot multi-speaker tts and zero-shot voice
  conversion for everyone.
\newblock In {\em International Conference on Machine Learning}, pages
  2709--2720. PMLR, 2022.

\bibitem{ye2024flashspeech}
Zhen Ye, Zeqian Ju, Haohe Liu, Xu~Tan, Jianyi Chen, Yiwen Lu, Peiwen Sun,
  Jiahao Pan, Weizhen Bian, Shulin He, et~al.
\newblock Flashspeech: Efficient zero-shot speech synthesis.
\newblock {\em arXiv preprint arXiv:2404.14700}, 2024.

\bibitem{sixmillion}
Devin Coldewey.
\newblock Six million fine for robocaller who used ai to clone biden’s voice.
\newblock
  https://techcrunch.com/2024/05/23/6m-fine-for-robocaller-who-used-ai-to-clone-bidens-voice,
  2024.
\newblock Online; accessed 29 May 2024.

\bibitem{rawnet2}
Hemlata Tak, Jose Patino, Massimiliano Todisco, Andreas Nautsch, Nicholas
  Evans, and Anthony Larcher.
\newblock End-to-end anti-spoofing with rawnet2.
\newblock In {\em ICASSP 2021-2021 IEEE International Conference on Acoustics,
  Speech and Signal Processing (ICASSP)}, pages 6369--6373. IEEE, 2021.

\bibitem{rawnetgatst}
Hemlata Tak, Jee-weon Jung, Jose Patino, Madhu Kamble, Massimiliano Todisco,
  and Nicholas Evans.
\newblock End-to-end spectro-temporal graph attention networks for speaker
  verification anti-spoofing and speech deepfake detection.
\newblock {\em arXiv preprint arXiv:2107.12710}, 2021.

\bibitem{aasist}
Jee-weon Jung, Hee-Soo Heo, Hemlata Tak, Hye-jin Shim, Joon~Son Chung, Bong-Jin
  Lee, Ha-Jin Yu, and Nicholas Evans.
\newblock Aasist: Audio anti-spoofing using integrated spectro-temporal graph
  attention networks.
\newblock In {\em ICASSP 2022-2022 IEEE international conference on acoustics,
  speech and signal processing (ICASSP)}, pages 6367--6371. IEEE, 2022.

\bibitem{muller2022does}
Nicolas~M M{\"u}ller, Pavel Czempin, Franziska Dieckmann, Adam Froghyar, and
  Konstantin B{\"o}ttinger.
\newblock Does audio deepfake detection generalize?
\newblock {\em arXiv preprint arXiv:2203.16263}, 2022.

\bibitem{xie2024codecfake}
Yuankun Xie, Yi~Lu, Ruibo Fu, Zhengqi Wen, Zhiyong Wang, Jianhua Tao, Xin Qi,
  Xiaopeng Wang, Yukun Liu, Haonan Cheng, et~al.
\newblock The codecfake dataset and countermeasures for the universally
  detection of deepfake audio.
\newblock {\em arXiv preprint arXiv:2405.04880}, 2024.

\bibitem{khan2016concatenative}
Rubeena~A Khan and Janardan~Shrawan Chitode.
\newblock Concatenative speech synthesis: A review.
\newblock {\em International Journal of Computer Applications}, 136(3):1--6,
  2016.

\bibitem{zen2009statistical}
Heiga Zen, Keiichi Tokuda, and Alan~W Black.
\newblock Statistical parametric speech synthesis.
\newblock {\em speech communication}, 51(11):1039--1064, 2009.

\bibitem{wang2017tacotron}
Yuxuan Wang, RJ~Skerry-Ryan, Daisy Stanton, Yonghui Wu, Ron~J Weiss, Navdeep
  Jaitly, Zongheng Yang, Ying Xiao, Zhifeng Chen, Samy Bengio, et~al.
\newblock Tacotron: Towards end-to-end speech synthesis.
\newblock {\em arXiv preprint arXiv:1703.10135}, 2017.

\bibitem{ping2017deep}
Wei Ping, Kainan Peng, Andrew Gibiansky, Sercan~O Arik, Ajay Kannan, Sharan
  Narang, Jonathan Raiman, and John Miller.
\newblock Deep voice 3: Scaling text-to-speech with convolutional sequence
  learning.
\newblock {\em arXiv preprint arXiv:1710.07654}, 2017.

\bibitem{ren2019fastspeech}
Yi~Ren, Yangjun Ruan, Xu~Tan, Tao Qin, Sheng Zhao, Zhou Zhao, and Tie-Yan Liu.
\newblock Fastspeech: Fast, robust and controllable text to speech.
\newblock {\em Advances in neural information processing systems}, 32, 2019.

\bibitem{ren2020fastspeech2}
Yi~Ren, Chenxu Hu, Xu~Tan, Tao Qin, Sheng Zhao, Zhou Zhao, and Tie-Yan Liu.
\newblock Fastspeech 2: Fast and high-quality end-to-end text to speech.
\newblock {\em arXiv preprint arXiv:2006.04558}, 2020.

\bibitem{kumar2019melgan}
Kundan Kumar, Rithesh Kumar, Thibault De~Boissiere, Lucas Gestin, Wei~Zhen
  Teoh, Jose Sotelo, Alexandre De~Brebisson, Yoshua Bengio, and Aaron~C
  Courville.
\newblock Melgan: Generative adversarial networks for conditional waveform
  synthesis.
\newblock {\em Advances in neural information processing systems}, 32, 2019.

\bibitem{doddington2001speaker}
George~R Doddington et~al.
\newblock Speaker recognition based on idiolectal differences between speakers.
\newblock In {\em Interspeech}, pages 2521--2524, 2001.

\bibitem{alegre2013new}
Federico Alegre, Ravichander Vipperla, Asmaa Amehraye, and Nicholas Evans.
\newblock A new speaker verification spoofing countermeasure based on local
  binary patterns.
\newblock In {\em INTERSPEECH 2013, 14th Annual Conference of the International
  Speech Communication Association, Lyon: France (2013)}, page~5p, 2013.

\bibitem{hanilcci2015classifiers}
Cemal Hanil{\c{c}}i, Tomi Kinnunen, Md~Sahidullah, and Aleksandr Sizov.
\newblock Classifiers for synthetic speech detection: A comparison.
\newblock 2015.

\bibitem{patel2015combining}
Tanvina~B Patel and Hemant~A Patil.
\newblock Combining evidences from mel cepstral, cochlear filter cepstral and
  instantaneous frequency features for detection of natural vs. spoofed speech.
\newblock In {\em Interspeech}, pages 2062--2066, 2015.

\bibitem{sahidullah2015comparison}
Md~Sahidullah, Tomi Kinnunen, and Cemal Hanil{\c{c}}i.
\newblock A comparison of features for synthetic speech detection.
\newblock 2015.

\bibitem{todisco2016new}
Massimiliano Todisco, H{\'e}ctor Delgado, and Nicholas~WD Evans.
\newblock A new feature for automatic speaker verification anti-spoofing:
  Constant q cepstral coefficients.
\newblock In {\em Odyssey}, volume 2016, pages 283--290, 2016.

\bibitem{rawnet}
Jee-weon Jung, Seung-bin Kim, Hye-jin Shim, Ju-ho Kim, and Ha-Jin Yu.
\newblock Improved rawnet with feature map scaling for text-independent speaker
  verification using raw waveforms.
\newblock {\em arXiv preprint arXiv:2004.00526}, 2020.

\bibitem{tak2021graph}
Hemlata Tak, Jee-weon Jung, Jose Patino, Massimiliano Todisco, and Nicholas
  Evans.
\newblock Graph attention networks for anti-spoofing.
\newblock {\em arXiv preprint arXiv:2104.03654}, 2021.

\bibitem{asvspoof2019}
Massimiliano Todisco, Xin Wang, Ville Vestman, Md~Sahidullah, H{\'e}ctor
  Delgado, Andreas Nautsch, Junichi Yamagishi, Nicholas Evans, Tomi Kinnunen,
  and Kong~Aik Lee.
\newblock Asvspoof 2019: Future horizons in spoofed and fake audio detection.
\newblock {\em arXiv preprint arXiv:1904.05441}, 2019.

\bibitem{frank2021wavefake}
Joel Frank and Lea Sch{\"o}nherr.
\newblock Wavefake: A data set to facilitate audio deepfake detection.
\newblock {\em arXiv preprint arXiv:2111.02813}, 2021.

\bibitem{visqol}
Andrew Hines, Jan Skoglund, Anil~C Kokaram, and Naomi Harte.
\newblock Visqol: an objective speech quality model.
\newblock {\em EURASIP Journal on Audio, Speech, and Music Processing},
  2015:1--18, 2015.

\bibitem{mkadry2017towards}
Aleksander Mkadry, Aleksandar Makelov, Ludwig Schmidt, Dimitris Tsipras, and
  Adrian Vladu.
\newblock Towards deep learning models resistant to adversarial attacks.
\newblock {\em stat}, 1050(9), 2017.

\bibitem{kurakin2018adversarial}
Alexey Kurakin, Ian~J Goodfellow, and Samy Bengio.
\newblock Adversarial examples in the physical world.
\newblock In {\em Artificial intelligence safety and security}, pages 99--112.
  2018.

\bibitem{simba}
Chuan Guo, Jacob Gardner, Yurong You, Andrew~Gordon Wilson, and Kilian
  Weinberger.
\newblock Simple black-box adversarial attacks.
\newblock In {\em International conference on machine learning}, pages
  2484--2493. PMLR, 2019.

\bibitem{liu2024audiomarkbench}
Hongbin Liu, Moyang Guo, Zhengyuan Jiang, Lun Wang, and Neil~Zhenqiang Gong.
\newblock Audiomarkbench: Benchmarking robustness of audio watermarking.
\newblock {\em arXiv preprint arXiv:2406.06979}, 2024.

\end{thebibliography}
\bibliographystyle{unsrt}
% \input{text/appendix}

% \appendix
% \section{Appendix}

\end{document}